\documentclass[preprint]{aastex}
\begin{document}

\title{{\it Chandra} Observations and Monte Carlo Simulations of the
Grain-Scattered Halo of the Binary X-Ray Pulsar 4U 1538-52}

\author{George W. Clark}
\affil{Center for Space Research, Massachusetts Institute of
        Technology, Cambridge, MA 02139}

\begin{abstract}
Properties of the grain-scattered X-ray halo of the eclipsing X-ray
binary pulsar 4U 1538-52 are derived from a 25 ksec observation by the
{\it Chandra} X-ray observatory extending from just before its eclipse
immersion to near mid-eclipse. Profiles of the observed halo, compiled
in two energy ranges, 2 to 4 keV and 4 to 6 keV, and three time
intervals before and after the eclipse, exhibit a three-peak shape
indicative of a concentration of the interstellar dust grains in three
discrete clouds along the line of sight.  The observed profiles are
fitted by the profiles of a simulated halo generated by a Monte Carlo
ray-tracing code operating on a model of three discrete clouds and a
spectrum of the photons emitted by the source over a period of time
extending from 270 ksec before the observation began till it ended. In
the model, the spectrum before the observation began is expressed as a
function of the orbital phase of the pulsar and is derived as an
average over 6.3 years of data accumulated by the All Sky Monitor of
the {\it Rossi X-Ray Timing Explorer}.  The distances of the two
nearer dust clouds are fixed at the distances of the peaks of atomic
hydrogen derived from the 21-cm spectrum in the direction of the X-ray
source, namely at 1.30 and 2.56 kpc. With these constraints, a good
fit is achieved with the source at a distance 4.5 kpc, the distance of
the third cloud at 4.05 kpc, the total scattering optical depth of the
three clouds equal to 0.159 at 3 keV, and the column density of
hydrogen set to $4.6\times 10^{22}$ cm$^{-2}$. With $A_{\rm
V}=6.5\pm0.3$~mag for the binary companion star, QV Nor, the ratio of
the scattering optical depth at 3 keV to the visual extinction is
$0.0234\pm0.0010$ mag$^{-1}$.  The column density of hydrogen in the
model is much greater than the column density of atomic hydrogen
derived from the 21-cm spectrum, which indicates that most of the
hydrogen is in molecules, probably concentrated in the three dust
clouds.
\end{abstract}

\keywords{dust, extinction---scattering---X-rays:binaries---pulsars: individual~(4U~1538-52)}

\section{INTRODUCTION}
The grain-scattered halos of X-ray sources contain clues to the nature
and distribution of interstellar dust grains.  The {\it Chandra} X-ray
observatory provides the means to achieve substantial improvements
over previous observations of X-ray halos.  Here we present results
from a {\it Chandra} observation and a Monte Carlo simulation of the
halo of the eclipsing X-ray binary pulsar 4U 1538-52 (hereinafter
x1538), an especially interesting halo because of its comparatively
high intensity, evidence in the shapes of its radial profiles for a
concentration of dust in discrete clouds along the line of sight, and
high variability caused by eclipses and flares.

The theory of grain-scattered halos of steady sources, first discussed
by \citet{ove65}, has been developed by various authors,
e.g. \citet{sly69,hay70,mar70,mau86,smi98,dra03}.  The effects
of intensity variations of a source on the variation of its halo and
the possibilities this can present for measuring the distance of the
source and the distribution of the scattering grains along the
sight line have been discussed by \citet{tru73,alc78,xu86,klo94}.
\citet{klo91} calculated the expected profiles of halos that would be
produced by grains with various assumed size distributions.
\citet{mat91} carried out an analytical study of the possibilities of
using halos as diagnostics of grains.  \citet{pre96} discussed the
possibilities and the practical limitations of such
diagnostics. \citet{smi98} showed that the Rayleigh-Gans approximation
to the Mie scattering theory, commonly employed in theoretical
treatments X-ray halos, yields substantial overestimates of halo
intensities for photon energies less than 1 keV.

Early evidence of X-ray halos was reported by \citet{rol83} and by
\citet{cat83} from images recorded by the {\it Einstein}
observatory. Mauche and collaborators \citep{mau86, mol86, mau89},
employing the same observatory, found a correlation between the
fractional intensity of the halo of an X-ray source and the visual
extinction of its optical counterpart that was consistent with the
model of solid silicate and graphite grains proposed by \citet{mat77}.
\citet{pre95} carried out an extensive survey of X-ray halos with the
{\it ROSAT} X-ray observatory. \citet{wit01} found evidence for a
distribution in grain size extending to 2 $\mu$m and possibly more in
a {\it ROSAT} observation of the halo of Nova Cygni
1992. \citet{pre00} derived an estimate of the distance to Cyg X-3 by
an analysis of its X-ray halo. \citet{cla94} employed the non-imaging
{\it Ginga} observatory to observe x1538 and its halo.  From a
comparison of the counting rates before and after an eclipse of x1538
they derived an estimate of the fractional halo intensity that was
substantially less than that expected on the assumption that the
visual extinction of its optical companion is caused by solid silicate
grains with a specific density near 3. They interpreted the result as
consistent with a grain structure in the form of loose aggregates of
smaller particles.

Following its discovery in the {\it Uhuru} galactic plane survey
\citep{gia74}, x1538 was observed to pulse with a period of 529 s and
to have an orbital period of 3.73 days with an eclipse lasting about
12 hours \citep{dav77,bec77}. An accurate position determination with
{\it SAS 3} \citep{app78} led to identification of the highly reddened
BO supergiant, QV Nor, as its companion \citep{cra78}. The X-ray
spectrum has a prominent iron K line \citep{mak87} and a pulse
phase-dependent cyclotron resonance feature at 20 keV
\citep{cla90}. Properties of the binary system derived from various
observations are listed in Table 1.

For the present study X-ray images of x1538 and its halo were recorded
before and after an eclipse immersion. Radial profiles of simulated
halos, generated by a Monte Carlo ray-tracing code, were fitted to the
three-peaked time-varying radial profiles of the observed halo by
adjustment of the parameters of a model of the source spectrum, the
dust grains, and the interstellar photoelectric absorption.  The model
parameters that yielded a good fit are compared with the optical
extinction of QV Nor and with the column density of atomic hydrogen
derived from the spectrum of 21-cm radiation in the direction of the
source to derive information about the scattering efficiency of the
grains and the total column density of hydrogen.

Section 2 describes the observations and data reduction.  Properties
of the halo are presented in section 3. The Monte Carlo simulation
code is described briefly in section 4 together with the model and the
fitting procedure.  Section 5 is a summary and discussion of results.
Section 6 lists conclusions drawn from this study.  

\section{OBSERVATIONS AND DATA PREPARATION}
Data were obtained from the imaging spectrometer (ACIS-I) at the focal
plane of the X-ray telescope of the {\it Chandra} observatory during a
continuous 25 ksec observation (Chandra archive OBS\_ID 90) beginning
2000 April 8.95 ($T_{0}=$MJD 51642.94881) at orbital phase 0.9185,
shortly before an eclipse immersion, and ending near mid-eclipse, at
orbital phase 0.001. The received data for each detected photon
consist of the transfer time, $T$, of its 3.200-s data frame, its
arrival direction expressed as X(EW),Y(SN) sky coordinates in units of
$0.492\arcsec$, and its energy with an uncertainty of about 0.1
keV.

Point response functions (PSFs) of the imaging system were derived
from a combination of results from a 30 ksec observation of the X-ray
bright quasar 3C 273 and a 12 ksec observation of the much fainter
quasar PG1634+70.

We prepared arrays of the effective area per second of exposure in
$3.94\arcsec \times 3.94\arcsec$ pixels as a function of sky position,
averaged over the duration of the observation and expressed in units
of cm$^{2}$ per second of observation, for eighteen energies between
0.5 and 8.0 keV.  In our tabulation of the received data for use in
the analysis we replaced the transfer time $T$ by $t=T-T_{0}$, and add the
reciprocal of the effective area per second of exposure (REA) as
determined according to its sky position and energy. 

The arrival directions of all events in the energy range from 2 to 6
keV from the approximate beginning of total eclipse at $t=5$ ksec to
near mid-eclipse at 25 ksec are plotted relative to the direction of
x1538 as dots in Fig. 1(a). The image of x1538 is in the northeast
corner of the region scanned by CCD chip I\,3. The halo image, which
persists with gradually diminishing intensity during the eclipse, is
conspicuous in a circular region spread over the four chips. Regions
of the sky scanned by CCD pixels near the edges of the chips have
relatively low effective areas achieved operationally by dithering the
telescope orientation in a pattern centered on a fixed direction with
a cycle time time of 16 s. We excluded from our analysis events in
these regions to avoid the large statistical fluctuations they would
otherwise cause in summations of REA.  We searched for other sources
in the field of view by mapping all the events recorded during the
interval from 1039 to 25,039 s in the energy range from 2 to 6 keV in
$2\arcsec \times 2\arcsec$ pixels.  We eliminated any significant
effect of other sources on the halo profiles by excluding events
within $2\arcsec$ of the centers of pixels containing 5 or more events
and more than 10$\arcsec$ from the image center.  Finally, we excluded
events within $2\arcsec$ of the center line of the ``transfer streak''
which is the trailed image of x1538 running parallel to the southeast
edge of the region scanned by chip 3 and formed during the 0.041 s
shift of the CCD data to the frame store for subsequent transmission.
Fig. 1(b) is a map of the events remaining after application of the
exclusion criteria.

To determine the effective solid angle on the sky corresponding to a
given region of the image we generated $10^{4}$ events with random
positions distributed uniformly over the region and computed the
fraction of events accepted according to the exclusion criteria.  The
effective solid angle was evaluated as the product of that fraction
times the geometrical solid angle of the region.  The incident
intensity of photons corresponding to a set of events recorded within
a specified region of the image and interval of time was computed as
the sum of the REAs of the events divided by the product of the
exposure time times the solid angle. The fractional error of an
intensity was estimated as the reciprocal square root of the number of
events.

Data from within $2\arcsec$ of the image center recorded before
the eclipse are severely degraded by pileup, i.e., multiple
occurrences of events in the same pixel during the 3.200-s frame
exposure. Under these circumstances the transfer streak is a
useful source of data on the source flux unaffected by pileup, albeit
with exposures smaller than the exposures of the frame data by the
product of the fraction of the exposure time occupied by transfers
(0.0128) times the fraction of the streak used.  To extract the streak
data, we summed the REA of events in a $4\arcsec$-wide rectangle
centered on the transfer streak and extending about half the length of
the streak. Background intensity was derived from events in parallel
rectangles between $3\arcsec$ and $10\arcsec$ from the center line of
the streak.

During the time before the eclipse, when the incident intensity of
unscattered photons from x1538 was highest, the average rate of all
events in the angle range from $12\arcsec$ to $13\arcsec$ was less
than $10^{-3}$ s$^{-1}$ arcsec$^{-2}$. Thus the effects of pileup on
measured photon intensities at angles greater than $12\arcsec$ was
negligible.  We derived halo profiles from events in the angle range
from $12\arcsec$ to $312\arcsec$ and preliminary estimates of the
background intensities from events in the angle range from
$480\arcsec$ to $600\arcsec$. Since the latter events included some
halo events, background intensities were included along with the PSF
normalizations among the parameters adjusted in the process of fitting
the profiles of the simulated halo to the observed profiles.
  
\section{PROPERTIES OF THE SOURCE AND HALO} 
\subsection{Spectra and Light Curves of the Source}
The pre-eclipse spectrum of the source is displayed in Fig. 2(a). It
was derived from the transfer streak during seven complete pulse
periods from 1037 to 4739 s.  The high value in the 6.0-6.5 keV
bin is due to the iron K line. The sharp decline below 4.0 keV is
caused by strong circumstellar absorption that occurs as the sight
line grazes the limb of QV Nor.  The histogram, derived from the
function
\begin{eqnarray}
s_{\rm pre}(E,t)=\left\{I_{0}E^{-1.12}+F_{\rm FE}
\frac{\rm keV}{\sqrt{2\pi}\sigma}\exp\left[-\frac{(E-E_{\rm
FE})^{2}}{2\sigma^{2}}\right]\right\} \exp[-N_{\rm H}\sigma_{\rm
pe}(E)-\tau_{\rm sca}(E/{\rm keV})^{-2}],\nonumber\\
\hspace{3in}1037 ~{\rm s}<t<4739 ~{\rm s},
\end{eqnarray}
was fitted by least squares to the data, with $F_{\rm Fe}=1.3\times
10^{-3} ~{\rm cm}^{-2} {\rm s}^{-1}$, $E_{\rm Fe}=6.4$ keV,
$\sigma=0.1$ keV, and the other parameter values listed in Table 2.
The quantity $\sigma_{\rm pe}(E)$ is the photoelectric cross section
per hydrogen atom of \citet{mor83}, and $N_{\rm H}$ is the total
column density of hydrogen along the sight line from the neutron
star. The optical depth for scattering is represented by $\tau_{\rm
sca}(E/{\rm keV})^{-2}$.

Fig. 2(b) displays the source spectrum during the eclipse,
derived from events within $2\arcsec$ of the image center from
5 to 25 ksec when pileup caused only a small reduction
in detection efficiency.  During the eclipse, X-rays in the image core
are fluorescent and Thomson-scattered radiation generated in distant
circumstellar matter without strong circumstellar absorption.  The
iron K radiation is especially prominent in the eclipse spectrum.

In our analysis and simulation of the halo we restricted consideration
to photons with energies in the range from 2 to 6 keV to exclude, on
the high end, the iron line photons and, on the low end, low-energy
photons for which the scattering cross section is less accurately
represented by the Rayleigh-Gans approximation used in the
simulations.

In the upper panel of Fig. 3 the incident flux of photons in the
energy range from 2 to 6 keV within $2\arcsec$ of the x1538 image
center is plotted as open circles against orbital phase (bottom
abscissa) and time from the start of the observation (top abscissa).
Solid circles show the flux derived from events in the transfer
streak.  In both cases integration times were set equal to the pulse
period of 529 s before the eclipse and to 4952 s after the eclipse
when pulsations were no longer present. The sharp drop in the streak
rate near orbital phase 0.933 indicates the eclipse edge
corresponding to an eclipse half-angle of $24.1\arcdeg\pm0.3\arcdeg$.
The drastically lower values of the open circles before the eclipse
are caused by pileup. After the eclipse, when X-rays scattered from
the circumstellar matter in the x1538/QV Nor system are still present
in the source image, the core pileup is much less severe and yields
values of greater statistical accuracy, though still systematically
less than the streak values.

\subsection{Properties of the Grain-Scattered Halo}
Light curves of the background-subtracted halo in two energy ranges, 2
to 4 keV and 4 to 6 keV, are shown in the lower panel of Fig. 3. The
initial decline of the halo flux at the eclipse is much less than that
of the direct source due to the continuing arrival after the eclipse
of photons delayed by scattering. The straight lines fitted by least
squares to the data from 5 to 25 ksec show the decay of the halo flux
during the eclipse.

Fig. 4 displays the spectra of the halo as average intensities in
successive intervals of angle from the source direction plotted
against energy.  The spectra were derived from background-subtracted
data recorded during the eclipse from 5 to 25 ksec. They exhibit
the progressive softening with increasing angle of scattering expected
from scattering theory.

Radial profiles of the halo, in two energy ranges and three time
intervals before and after the eclipse, are displayed in Fig. 5.  The
solid circles represent values of the incident intensity of photons in
thirty $10\arcsec$ intervals of angle from $12\arcsec$ to $312\arcsec$
from the source, with backgrounds and PSFs of the source subtracted.
The error bars indicate the Poisson counting uncertainties of the raw
data.  

The profiles appear to have a three-peak shape.  The outer two peaks
are clearly defined by the low values of the intensities in bins 5 and
10 of panel (3) and bins 6 and 13 of panel (5).  The low values in bin
1 of all three of the 2-4 keV profiles, though of low statistical
significance, indicate the possible presence of a third peak.  A
uniform distribution of dust gives rise to a halo the profile of
which, after an eclipse, develops a single broad peak whose inner
bound moves toward larger angles as time from eclipse ingress
increases.  To explain the formation of multiple peaks, it therefore
appears necessary to assume that the dust is not uniformly
distributed. The problem we tackled was to find a model of the
non-uniform grain distribution and the variable source spectrum that
can account for the profiles of a halo produced by multiple scattering
near the line of sight.  We developed a Monte Carlo ray-tracing code
to circumvent the difficulties of deriving halo profiles from such a
model by analytical computation.
   
\section{MONTE CARLO GENERATION OF SIMULATED HALOS} 
The code generates the energies, arrival directions, and arrival times
of contributions to a halo from scatterings of trial photons emitted
by a source with a specified variable spectrum and scattered by dust
grains with a specified distribution in size and a specified
distribution of the optical depth for scattering along the sight line.
The code launches a specified number of trial photons from the source
each second beginning far enough in the past to assure that all
significant contributions to the halo during the observation are
included, and ending at the termination of the observation.  Each
trial photon has an energy selected at random from the spectrum.  It
is tracked in three dimensions till it passes beyond the observer or
has accumulated so great a delay in transit as to arrive after the end
of the observation.

The trajectory of a trial photon is described in a rectangular
coordinate system with its origin at the source, and the Z axis along
the sight line to the observer. Distance from the source along the Z
axis is expressed as a fraction, $z$, of the total distance, $D$, from
source to observer. The delay in the arrival time of a contribution to
the halo intensity from a given scattering, relative to the arrival
time of an unscattered photon, is the sum of the delays in successive
path segments each of which is computed in the small angle
approximation as $(z_{k}-z_{k-1})D\psi_{k}^{2}/2c$, where $z_{k}$ is
the projected position on the Z axis of the $k$th scattering,
$\psi_{k}$ is the angle that the $k$th segment makes with the Z axis,
and $c$ is the speed of light. The contribution to the halo intensity
from each scattering is tabulated according to its energy, arrival
angle, arrival time, the identification number of the dust cloud in
which the scattering occurred, and the number of the scattering.

We assume the photoelectric absorption and grain scattering depend
only on the Z component of position.  At each scattering the size of
the grain is randomly selected from a specified size distribution. We
assume the probability per steradian that a photon of energy $E$
scattered by a grain of size $a$ will be deflected at an angle $\phi$
varies with size and energy according to the relation
\begin{equation}
\frac{{\rm d}P}{{\rm d}\Omega}\sim a^{6}\left[\frac{F(E)}{Z}\right]^{2}\Phi(u)^{2},
\end{equation}
where $\Phi(u)$ is the form factor for homogeneous spherical particles
of radius $a$, and
\begin{eqnarray}
u=\frac{4\pi Ea}{hc}\sin\frac{\phi}{2},
\end{eqnarray}
where $h$ is Planck's constant.  For $E> 2$ keV and $a\leq 0.3 \,\mu
{\rm m}$ we can, with good accuracy, set the atomic scattering factor
$F(E)/Z=1$ \citep{hen81}, and represent the form factor by the
Rayleigh-Gans approximation \citep{van81},
\begin{equation}
\Phi(u)=3\frac{\sin u-u\cos u}{u^{3}}.
\end{equation}

The code operates on a model that specifies
\begin{itemize}
\item the distance of the source; 
\item the time-dependent spectrum of X-ray photons coming directly
from the x1538/QV Nor system before and after the observation begins;
\item the distribution in size of dust grains;
\item the scattering optical depth measured from the source along the
line of sight as a function of $z$;
\item the optical depth for photoelectric absorption;
\item the background intensity;
\item the PSF of the imaging system.
\end{itemize}

At the conclusion of the computation the accumulated data are analyzed
for the properties of the simulated halo in a manner similar to that
employed with real data, but with the added feature of having
available for each contribution the number of the scattering and the
identification of the cloud from which it came.  In addition, we sum
the contributions in intervals of the launch times to check whether
the specified beginning launch time is sufficiently early to capture
all significant contributions to the simulated halo.

Statistical fluctuations in Monte Carlo simulations and strong
cross-correlations in the effects of changes in the model parameters
preclude effective use of the code in an automatic curve-fitting
procedure.  Instead, we sought a fit of simulated profiles to observed
profiles in multiple trials with adjustments of the model parameters
aimed at reducing the mean square deviation between the observed and
simulated profile intensities.

Figure 14 shows a simulated halo profile that can be compared with an
analytical solution computed by \citet{mat91} for a steady source of 1
keV photons scattered by 1 $\mu {\rm m}$ grains distributed uniformly
along the sight line with a scattering optical depth of 2. The
simulated and analytical profiles of the halo components from first
and second scatterings are in close agreement. Small systematic
differences appear in the third and fourth scattering components which
may be due to a geometrical approximation in the analytical solution,
namely, the use at each scattering beyond the first of a ``spherical
coordinate system at the grain with a polar angle $\theta^{\prime}$
measured from the point source'' instead of from the previous
scattering site. Figure 15 shows the evolution of this model halo that
would occur following an eclipse immersion at 20 ksec and emersion at
60 ksec.

\subsection{Specification of the model}
In many trials, good fits of simulated profiles to the observed
profiles were found with widely different sets of model
parameters. Effects of a change in a given parameter can often be
reversed by a compensating change in another.  For example, increase
in halo intensity caused by an increase in the source flux can be
approximately cancelled by a decrease in the scattering optical depth
that leaves the product of the two parameters unchanged.  Change in
the angular position of a feature in the halo of a variable source
caused by a decrease in the assumed distance, $d$, of a scattering
site can be offset by a decrease in the assumed distance $D$ of the
source that leaves the quantity $(1/d-1/D)$ unchanged.  Therefore, to
obtain results that might provide useful information about the
distance of the source and the properties and distribution of the
dust, it was necessary to fix as many of the parameters as possible at
plausible values derived from this and other observations.

Previous studies show that the spectrum of x1538 is highly variable
during a single binary orbit.  To allow for this, we represent the
incident spectrum at time $t$ and orbital phase $\Pi(t)$ by
\begin{equation}
s_{\rm inc}(E,t)=s_{\rm av}(E,\Pi)[1+f(t)]+s_{\rm pre}(E,t)+s_{\rm
ecl}(E),
\end{equation}
in which $s_{\rm av}(E,\Pi)$ is the long-term average incident
spectrum of photons that arrive directly from the neutron star,
$[1+f(t)]$ is a factor representing flares that may have occurred
before the observation began, $s_{\rm pre}(E,t)$ is the observed
pre-eclipse spectrum of x1538 derived from the transfer streak, and
$s_{\rm ecl}(E)$ is the spectrum of photons observed during the
eclipse.  We assume the latter, characterizing photons produced by
X-ray illumination of the circumstellar matter in the x1538/QV Nor
system, is constant throughout the binary orbit.

We represent the average spectrum by
\begin{equation}
s_{\rm av}(E,\Pi)=I_{0}E^{-\alpha} \exp[-N_{\rm
H}(\Pi)\sigma_{\rm pe}(E)-\tau_{\rm sca}(E/{\rm keV})^{-2}].
\end{equation}
The quantity $N_{\rm H}(\Pi)$ is the column density of hydrogen between
x1538 and the observer, and $\sigma_{\rm pe}(E)$ is the photoelectric
cross section per hydrogen atom. We assume $N_{\rm H}(\Pi)$
is the sum of a constant interstellar component, $N_{\rm H}^{\rm is}$, and
a variable circumstellar component,
\begin{eqnarray}
N^{\rm cs}_{\rm H}(\Pi)&=&N_{\rm ecl}\exp[-(\Pi-\Pi_{\rm ecl})/\lambda_{\rm out}],	\hspace{3.0cm} \Pi<0.5,\\
			&=&N_{\rm ecl}\exp[-(1-\Pi-\Pi_{\rm ecl})/\lambda_{\rm in}),
	\hspace{2.4cm} \Pi>0.5,
\end{eqnarray}
where $\Pi_{\rm ecl}$ is the orbital phase of total eclipse, and
$N_{\rm ecl}$ is the column density as the sight line grazed the edge
of QV Nor. 

The average spectrum was derived from data in the archive of the All
Sky Monitor (ASM) of the {\it Rossi X-Ray Timing Explorer} (RXTE). The
archive provides a background-subtracted aspect-corrected count rate
for a given source whenever it is in the field of view of one of three
detectors during an orientation dwell in the course of its all-sky
scan.  We used the 16752 B-channel (nominal 3-5 keV) count rates in
the range from -3.0 to 3.0 s$^{-1}$ of detectors 1 and 2 during the
time interval from MJD 50088 to 52400 when the corresponding rates for
the Crab Nebula were steady with an average of 23.08 s$^{-1}$. The
data were folded in 50 intervals of orbital phase derived from the
orbital parameters listed in Table 1. The average rates are plotted
against orbital phase in Fig. 6(a).  The smooth curve is the function
\begin{equation}
C_{3-5}^{i}=\int_{\Pi_{i-1}}^{\Pi_{i}}d\Pi\int_{3 \,\rm keV}^{5\,
\rm keV}s_{\rm av}(E,\Pi)A(E)dE+C_{0},\hspace{.5cm}i=1...50.
\end{equation}
We set $N_{\rm H}^{\rm is}$ and $\tau_{\rm sca}$ to the values derived
in fitting the simulated profiles to the observed profiles.  The phase
of total eclipse, $\Pi_{\rm ecl}$, was set to the value indicated by
the vertical line in Fig. 3. The factor $A(E)$ is the effective area
of the ASM detectors normalized to yield a predicted 3-5 keV count
rate for the Crab Nebula equal to the observed rate under the
assumption that the Crab spectrum is $9.5(E/{\rm keV})^{-2.1}\exp[-3.1\times
10^{21}\sigma_{\rm pe}(E)]$ cm$^{-2}$ s$^{-1}$ keV$^{-1}$.  The
function was fitted by least squares to the ASM data by adjustment of
$I_{0}$, $\lambda _{\rm out}$, $\lambda_{\rm in}$, and $F_{0}$.  The
quantity $C_{0}$, representing the counting rate in eclipse, is a
measure of the sum of rates due to photons scattered in the
circumstellar matter of the x1538/QV Nor system, a portion of the
x1538 halo, other sources in the field of view, and possible errors in
the background subtraction.  As such, its presence in equation (9) makes
the fitted value of $I_{0}E^{-\alpha}$ a good measure of the average
spectrum of photons that would be incident in the absence of
scattering and photoelectric absorption. We ignore the small error due
to halo contamination that remains because the halo contribution to
$C_{0}$ decays during the eclipse.  The fixed and fitted values of the
average spectrum parameters are listed in Table 3.

The observed flux of x1538 before the eclipse in the interval from
1037 to 4739 s is more than an order of magnitude larger than the
value implied by the average spectrum fitted to the ASM data.  The
column density of the spectrum fitted to the pre-eclipse streak data,
though large, is much less than that of the ASM average spectrum in
the same orbital phase range.  Thus it appears that there may have
been, in effect, a kind of hole in the circumstellar matter that
allowed an attenuated flux from x1538 to shine through. To model this
portion of the pre-eclipse spectrum, which has a substantial effect on
the simulated halos at small angles, we added the fitted streak
spectrum previously described, $s_{\rm pre}(E,t)$, to the model
spectrum in the phase interval from 0.922 to 0.933.

We set $s_{\rm ecl}(E)=5.0\times 10^{-5}$ cm$^{-2}$ s$^{-1}$
keV$^{-1}$ to represent the observed source spectrum during the eclipse
displayed in Fig. 2(b), and which is nearly flat in the energy range
from 2 to 6 keV.  It made only a very small contribution to the
simulated halos.

Characteristic flare phenomena of x1538 are shown in Fig. 6(b) which
is a plot against orbital phase of the average counting rates of x1538
in intervals of the pulse period recorded by the Proportional Counter
Array (PCA) of the RXTE in a pointed-mode observation
\citep{cla00}. During a flare the count rates can rise above average
by a factor of five or more for several pulse periods. Photons from
such a flare, scattering from dust in a discrete cloud, can produce a
sharp feature in the halo profile which moves to larger angles as the
time interval between the flare occurrence and the observation
increases.  To add a $k$th flare to the spectrum model we specified a
constant value, $f_{k}$, of $f(\Pi)$ beginning at a certain time
$\Pi_{k}$ and lasting for an orbital phase interval of 0.01 which
corresponds to a time interval of 3222 s.

With the incident spectrum before and during the observation
specified, the spectrum that would be incident in the absence of
interstellar photoelectric absorption and scattering is related to the
incident spectrum by the equation
\begin{equation}
s_{\rm src}(E,t)=s_{\rm inc}(E,t)\exp[N_{\rm H}^{\rm is}\sigma(E)+\tau_{\rm
sca}(E/{\rm keV})^{-2}].
\end{equation}
In simulations the energy of a trial photon is selected at random from
this source spectrum.

The distribution in size of the grains was specified by the power law
\citep{mat77}
\begin{equation}
\frac{dn}{da}\sim a^{-3.5}, \hspace{.5in} a_{\rm min}<a<a_{\rm max}.
\end{equation}

We call $\tau(z)(E/1~{\rm keV})^{-2}$ the optical depth from the
source to $z$ for scattering photons of energy $E$, and note that
$\tau(1)=\tau_{\rm sca}$. For a clue to the distribution of the dust
grains along the sight line we examined the spectrum of 21-cm
radiation in the direction of x1538 (l$^{\rm II}$=327\fdg4\,,\,b$^{\rm
II}$=2\fdg1), kindly provided by N. McClure-Griffiths from the
Southern Galactic Plane Survey \citep{mcc01} and displayed in
Fig. 7(a).  The sight line is sufficiently inclined with respect to
the galactic plane so that we can safely assume most of the hydrogen
along it lies between Earth and the tangent point at 6.7 kpc.
Assuming $R_{0}=8.0$ kpc and $R\Omega(R)=220$ km s$^{-1}$, we
converted the Doppler velocities to radial distance to obtain the
brightness temperature as a function of distance, $d$, as plotted in
Fig. 7(b).

We explored the viability of a model with dust density simply
proportional to the brightness temperature, i.e., $\tau(d/D)\propto
T_{\rm b}(d)$, and the source spectrum described above without
flares. Fig. 8 illustrates the gross failure of such a model to yield
a simulated halo with a profile that conforms to the peaks and valleys
of the observed profiles.

A halo due to scattering from a uniform distribution of dust decays
slowly outward from the center, giving rise to a profile with a peak
that moves to larger angles as time from eclipse ingress increases.
If the dust is concentrated in a narrow cloud, the hollowing out
occurs more rapidly. The rate at which the peak moves toward larger
angles depends on the location of the cloud and the distance of the
source. With this in mind, and with the intent to produce multi-peaked
profiles with a model of dust concentrated in several narrow clouds,
we call $\tau_{k}(E/1~{\rm keV})^{-2}$ the scattering optical depth of
the $k$th cloud at distance $d_{k}$ for photons of energy $E$ with
$\tau_{k}$ distributed uniformly from $z_{k}=1.-d_{k}/D$ to
$z_{k}+0.01$. With all the grains located in such clouds, $\tau_{\rm
sca}=\Sigma \tau_{k}$.

We placed two clouds at the distances of the two prominent peaks of
atomic hydrogen density at $d_{1}=1.30$ kpc and $d_{2}=2.56$ kp, and
explored how well the resulting two peaks in the profile intensity
could be made to conform to the outer two peaks of the observed
profiles by adjustment of the source distance and the cloud
thicknesses.  With $d_{1}=1.30$ kpc and $d_{2}=2.56$ kpc, and with
$\tau_{1}$ and $\tau_{2}$ adjustable, we generated halos for source
distances of 4.0, 4.5, 5.0, and 6.0 kpc, all within the wide range of
uncertainties of published estimates based on photometric measurements
of QV Nor. As in the previous example, we assumed the source spectrum
was as described above without flares. The results, displayed in
Fig. 9, show that the peaks shift, as expected, to larger angles as
the source distance is increased.  For distances of 5 and 6 kpc the
simulated profiles are poor matches to the observed profiles. The fit
for 4.5 appears slightly better than for 4.0 kpc.  Though 4.5 kpc is
near the lower limit of photometrically plausible distances, we
adopted it for the model, with the expectation that the poor fit at
angles below $60\arcsec$ could be repaired by placing a third cloud
closer to the source.

\subsection{Fitting Simulated Profiles to the Observed Profiles}
The six profiles displayed as solid circles with error bars in Fig.
11 were the targets of the fitting process.  They were compiled from
the same data as were the six profiles in Fig. 5, but without
subtraction of the background and the point spread functions.  To
achieve a fit, we added a third cloud to the model, and reduced the
rms deviation between the simulated and observed profiles as much as
seemed possible by adjusting $a_{\rm max}, \tau_{1}$, $\tau_{2}$,
$d_{3}$, $\tau_{3}$, and $N_{\rm H}^{\rm is}$, together with the
background levels, and the normalization factors of the point spread
functions. As expected, we could increase the intensity of a given
profile peak by increasing the scattering optical depth of the
corresponding cloud, but with correlated effects on the other peaks
due to shifts in the relative frequencies of scattering in the other
two clouds. We could increase the ratio between the intensities in the
4-6 and 2-4 keV profiles by increasing $N_{\rm H}^{\rm is}$.

We then sought values of the occurrence times and magnitudes of flares
that could cause single-bin deviations to mimic those that seemed to
be present, for example, in bin 18 in panel 3 and bin 15 in panel
5. The choice of occurrence times was guided by the equation
\begin{equation}
\theta=\left[2c(t-t^{\prime})(\frac{1}{d}-\frac{1}{D})\right]^{1/2},
\end{equation}
which expresses the arrival angle of a photon emitted at time
$t^{\prime}-D/c$ from the source which arrives at time $t$ after
scattering from a grain at distance $d$ from the observer. We note
that with three clouds the effects of one flare appear around three
separate angles.  Thus adding a flare to the model had subtle effects
that reduced the overall rms deviation of the fit, but are not all
evident to the eye.

Table 4 lists the fixed and fitted parameters of the final model, and
Fig. 10 displays some of the control data in graphical form.  The
intensities of the resulting simulated profiles are plotted as the
open circles in Fig. 11.  The curves indicate the sums of the
backgrounds and point spread functions that were added to the
simulated profiles.  Plotted below the profiles are the deviations of
the simulated data from the observed data divided by the one-sigma
statistical errors of the observed data.  The quality of the fit for
each profile is indicated in the plot by the rms deviation. The rms
deviation of the total fit is 0.88.  The statistical uncertainties of
the simulated data were negligibly small by virtue of the large number
(10$^{4}$) of trial photons launched per second, which yielded $\sim
3.4\times 10^{7}$ contributions to the simulated data.

The separate contributions to the simulated profiles by scatterings
from the three dust clouds are plotted in Fig. 12.
 	
Another test of the simulation is provided by the comparison between
the spectra of the observed and simulated halos displayed in Fig.
13. Both the observed and simulated spectra were compiled from data in
the angle range from $30\arcsec$ to $200\arcsec$ and the time range from
5037 to 25037 s.

Although a good fit has been obtained with the listed parameters, a
substantial uncertainty must be attached to each of them. A major
source of that uncertainty is the use of the ASM average spectrum for
lack of specific information about the spectrum during the period of
$\sim$2 days before the observation began.  Another source of
uncertainty are the cross correlations among the effects of variations
of the various model parameters which result in a very flat chi-square
function of the overall fit. We found that a good fit can be achieved
with the source distance fixed at 6 kpc, provided the distances of the
two nearer dust clouds are changed to 1.42 and 3.20 kpc, well away
from the distances of the atomic hydrogen concentrations derived from
the 21-cm spectrum. Fixing the distances of the two principal dust
clouds did place a significant constraint on the choice of the
adjusted parameters. Nevertheless, lacking an objective procedure for
evaluating the uncertainties in the fitted parameters that arise from
these ambiguities, we have listed the values that yielded the fit
without errors.

\section{SUMMARY OF RESULTS AND DISCUSSION}
The profiles of the grain-scattered halo of x1538, recorded before and
after an eclipse, have been fitted by the profiles of a simulated halo
generated by a Monte Carlo ray-tracing program operating on a model of
the dust grains and the source spectrum.  In the model the dust is
concentrated in three discrete clouds of which the two responsible for
most of the scattering are fixed at the distances of two peaks in the
density of atomic hydrogen derived from the 21-cm spectrum, namely
1.30 and 2.56 kpc. The source spectrum before the observation began
was assumed to be an average derived from several years of
observations by the All Sky Monitor on the {it RXTE} X-ray
observatory. The essential fitted parameters of the model were the
distance of x1538, the distance of the third cloud, the scattering
optical depths of all three clouds, and the column density of
hydrogen. Three flares were added to the model spectrum before the
observation began with occurrence times and magnitudes adjusted to
mimic sharp features in the observed profiles.

The model parameters of general significance are the source distance,
the total scattering optical depth, the interstellar photoelectric
absorption, and the upper limit on the grain size.

Locating the two principal dust clouds at the peaks in the density of
atomic hydrogen forced a choice of the source distance in the fitting
model that is substantially less than the estimates of 6.0 kpc and 6.4
kpc derived from photometry of QV Nor, respectively, by \citet{ilo79}
and \citet{rey92}.  The adopted value of 4.5 kpc is in the low end of
the uncertainty range of the estimate obtained from photometry by
\citet{cra78}.

The ratio
\begin{equation}
R_{\rm xv}(E)=\tau_{\rm sca}(E/{\rm keV})^{-2}/A_{\rm V}
\end{equation} 
is a measure of the scattering efficiency of the grains for photons of
energy $E$.  In this ratio, $A_{\rm V}$ is the visual extinction of
the optical companion, QV Nor. In the fitted model, $\tau_{\rm
sca}(E/{\rm keV})^{-2}=0.159$ for $E=9$ keV.  Assuming $A_{\rm
V}=6.8\pm0.3$ mag for the visual extinction of QV Nor, we find $R_{\rm
xv}(3~{\rm keV})=0.0234\pm0.0010$ mag$^{-1}$, where the stated error
includes only the uncertainty range of the published values of the
visual extinction. This result contradicts the much lower value
derived from the {\it Ginga} observation of x1538 previously cited
\citep{cla94}. However, it is somewhat lower than the theoretical
value of 0.0272 read from the plot computed by \citet{dra03} for solid
spherical carbonaceous and silicate grains, and is consistent with
the idea that interstellar dust grains are not compact.

The value of $N_{\rm H}$ required to match the ratio of the 2-4 keV
halo intensities to the 4-6 keV intensities, like the values obtained
from other observations of x1538, is much larger than the integrated
column density of atomic hydrogen derived from the 21-cm Doppler
spectrum in the direction of x1538.  The latter, computed according to
the equation \citep{bur02}
\begin{equation}
N_{\rm H}=1.8\times10^{18}\int T_{\rm b}{\rm d}v~{\rm cm}^{-2}, 
\end{equation}
is plotted against radial distance in Fig. 7(c).  If the photoelectric
cross section per hydrogen atom we have used is accurate, then most of
the hydrogen along the sight line to x1538 must be in molecules,
presumably in the three dust clouds.

\section{CONCLUSIONS}
From this study we conclude:
 
1) Monte Carlo simulation of a grain-scattered X-ray halo can provide
an effective means for extracting information about the distance of
the source and the properties and distribution of interstellar dust
grains in circumstances involving discrete dust clouds, source
variability, and multiple scattering.

2) To reduce uncertainty in such a simulation it is essential to have
an accurate record of the source spectrum for an extended period
before the halo observation begins.

3) The halo of x1538 is an especially interesting object because of
its relatively high surface brightness, complex structure, and high
variability due to eclipses and flares.

4) The time-varying three-peaked halo profiles of x1538 before and
after an eclipse can be fitted by simulated halo profiles generated by
a Monte Carlo ray-tracing code operating on a model with the source at
4.5 kpc, three discrete dust clouds of which two are at the distances
of the two greatest concentrations of atomic hydrogen derived from the
21-cm spectrum, and a third cloud closer to the source.

5) The total scattering optical depth of the three dust clouds in the
model is 0.159 for $E=3$ keV.  The ratio of this quantity to the visual
extinction of the optical companion, QV Nor, is $0.0234\pm0.0010$~mag$^{-1}$,
somewhat less than the theoretical value of 0.0272 derived for solid spherical
carbonaceous and silicate grains by \citet{dra03}.

\acknowledgments 

I thank Mark Bautz, John Arabadjis and the Chandra support staff for
help in use of the {\it Chandra} data system, and Ronald Remillard,
Allan Levine, and Edward Morgen for advice in the interpretation of
the RXTE data. I thank Naomi McClure-Griffiths for providing the 21-cm
spectrum in the direction of x1538.

\clearpage
\begin{deluxetable}{llll}
\tablewidth{5.0in} \tablecaption{Parameters of the 4U~1538-52/QV Nor
Binary System} \tablehead{\colhead{4U 1538-52~(ref. 5)}&&\colhead{QV
Nor}&\colhead{ref.}}  
\startdata 
Epoch=MJD 50449.93400 &		&D=$5.5\pm1.5$ kpc 		      &1\\ 
$a_{\rm x}\sin i=56.6\pm0.7$~lt-s &&\hspace{.11in}=$6.0\pm 0.5$ kpc   &2\\ 
$T_{\pi/2}=$MJD~$ 50450.206\pm0.014$&&\hspace{.11in}=$6.4\pm 1.0$ kpc &4\\ 
$P_{\rm pulse}=528.809\pm0.008$~s&& $A_{\rm V}=6.5$~mag 	      &1\\ 
$\dot\nu_{\rm pulse}=(-0.4\pm2.3)\times10^{-13}$~Hz s$^{-1}$& 
				&\hspace{.25in}=7.1 mag		      &2\\
$e=0.174\pm0.015$ &		&\hspace{.25in}=6.9 mag 	      &3\\
$\omega=64^{\circ}\pm9^{\circ}$&&				      & \\
$a_{0}$\tablenotemark{a}$~=47221.488\pm 0.015$~d &&		      & \\
$a_{1}$\tablenotemark{a}$~=3.728366\pm 0.000032$~d&&		      & \\
$a_{2}$\tablenotemark{a}$~=(-5.5\pm 4.0)\times 10^{-8}$~d &&	      & \\
$P_{\rm orb}=a_{1}+2a_{2}N$ &&					      & \\ 
$\dot P_{\rm orb}/P_{\rm orb}=(-2.9\pm2.1)\times10^{-6}$~yr$^{-1}$&&  & \\

\enddata \tablenotetext{a}{$T_{\pi/2}(N)=a_{0}+a_{1}N+a_{2}N^{2}$}
\tablerefs{(1)\citet{cra78}; \newline(2) \citet{ilo79}; (3) \citet{pak83};\newline(4)\citet{rey92}; (5) \citet{cla00}}
\end{deluxetable}

\clearpage
\begin{deluxetable}{lllllll}
\tablewidth{6.0in} \tablecaption{Spectrum parameters of x1538 in the
energy range less than 6 keV from various satellite observations.}

\tablehead{ \colhead{satellite} & \colhead{$I_{0}(\times 10^{-2})$} &
\colhead{$\alpha$} & \colhead{$N_{\rm H}(\times 10^{22})$} &
\colhead{$\tau_{\rm sca}$} & \colhead{orbital} &
\colhead{ref.}\\
 
\colhead{} & \colhead{cm$^{-2}$ s$^{-1}$
keV$^{-1}$} & \colhead{} & \colhead{cm$^{-2}$} & \colhead{} &
\colhead{phase} & \colhead{}}
 
\startdata
 
{\it Tenma}
&$3.7 \pm 0.3$ &$1.12\pm 0.04$ & $3.7\pm 0.4$ & 0.0 &high state      &1\\
 
{\it Ginga} & $4.87\pm 0.14$ & $1.19\pm 0.02$ & $2.0$ & 0.0&0.28-0.45&2\\
 
{\it SAX} & $6.6\pm 0.4$ & 1.12 & $1.80\pm 0.1$ & 0.0 &0.50-0.65&3\\
	  & $4.3\pm 0.2$ & 0.97 & $1.40\pm 0.1$ & 0.0 &0.65-0.75&3\\
{\it Chandra} & $3.4\pm 0.5$ & 1.12 & $16.5\pm0.5$ & 1.42 &0.921-0.933 & 4\\
\enddata 

\tablerefs{(1) \citet{mak87}; (2) \citet{cla94}; (3) \citet{rob01}; (4)
present data.}
 
\tablecomments{The spectra have the form\\ $I(E)=I_{0}\exp[-N_{\rm
H}\sigma(E)-\tau_{\rm sca}(E/{\rm keV})^{-2}](E/{\rm keV})^{-\alpha}$.}

\end{deluxetable}

\clearpage
\begin{deluxetable}{ll}
\tablewidth{4.5in}
\tablecaption{Parameters of the function fitted to the folded light
curve of x1538 compiled from the RXTE-ASM data  archive.}
\tablehead{\colhead{FIXED}&\colhead{FITTED}}
\startdata
$\Pi_{\rm ecl}=0.0675$
	&$I_{0}=4.03\times 10^{-2}$ cm$^{-2}$ s$^{-1}$ keV$^{-1}$	\\
$N_{\rm H}^{\rm ecl}=3.0\times 10^{25}$ cm$^{-2}$&
 	$\lambda_{\rm out}=0.263$ 					\\
$N_{\rm H}^{\rm is}=4.6\times 10^{22}$ cm$^{-2}$ &
	$\lambda_{\rm in}=0.258$					\\
$\tau_{\rm scat}=1.431$&
	$C_{0}=3.94\times 10^{-2}$ s$^{-1}$				\\
$\alpha=1.12$								\\ 
\enddata
\end{deluxetable}

\clearpage
\begin{deluxetable}{ll}
\tablewidth{4.3in}
\tablecaption{Fixed and fitted parameters of the halo simulation model.}
\tablehead{\colhead{FIXED}&\colhead{FITTED}}
\startdata
$I_{\rm av}=4.03\times 10^{-2}$ cm$^{-2}$ s$^{-1}$ keV$^{-1}$
				&$D=4.5$ kpc    			\\	
$a_{\rm min}=0.0003 ~\mu$m	
		&$N_{\rm H}^{\rm is}=4.6\times 10^{22}$ cm$^{-2}$	\\
$\alpha=1.12$ 			&$a_{\rm max}=0.27~\mu$m		\\
$\lambda_{\rm out}=0.263$	&$\tau_{1}=0.962$			\\
$\lambda_{\rm in}=0.258$	&$\tau_{2}=0.321$			\\
$\Pi_{\rm ecl}=0.0675$		&$d_{3}=4.05$~ kpc			\\
$N_{\rm H}^{\rm ecl}=3.00\times 10^{25}$ cm$^{-2}$
				&$\tau_{3}=0.148$			\\
$I_{\rm pre}=3.49\times10^{-2}$ cm$^{-2}$ s$^{-1}$ keV$^{-1}$
				&$\tau_{\rm sca}=1.431$			\\
$N_{\rm H}^{\rm pre}=16.5\times 10^{22}$ cm$^{-2}$
				&$t_{1}=-9.12$ ksec 			\\
$I_{\rm ecl}=5.00\times10^{-5}$ cm$^{-2}$ s$^{-1}$ keV$^{-1}$
				&$f_{1}=6.0$				\\
$d_{1}=1.30$~ kpc		&$t_{2}=-67.18$	ksec			\\
$d_{2}=2.56$~ kpc		&$f_{2}=1.5$				\\
				&$t_{3}=-115.51$ ksec			\\
				&$f_{3}=1.7$				\\
\enddata
\end{deluxetable}

\begin{figure}
\plotone{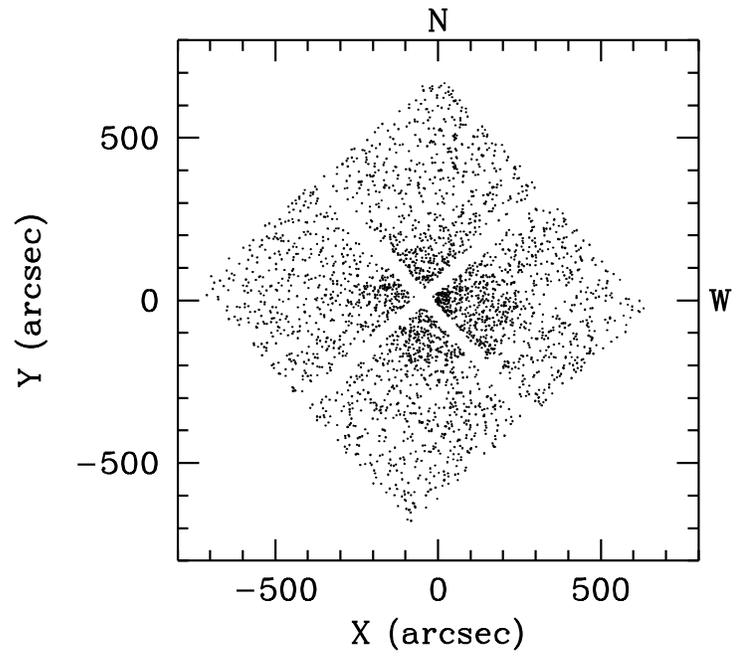} 
\figcaption[fig1.ps]{Map of events from the observation of
x1538 recorded after application of rejection
criteria.}
\end{figure}

\clearpage
\begin{figure}
\plotone{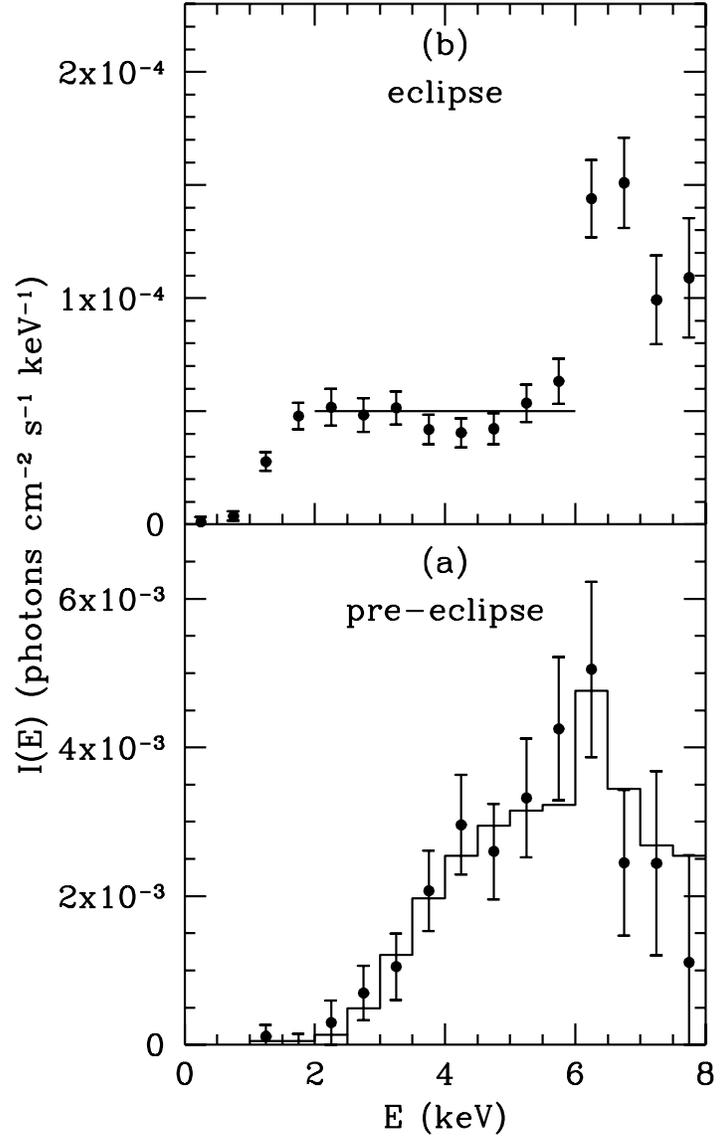} \figcaption[fig2.ps]{(a) Pre-eclipse spectrum of
x1538 derived from the transfer streak. (b) Post-eclipse spectrum
derived from events within $2\arcsec$ of the image center.}
\end{figure}

\clearpage
\begin{figure}
\plotone{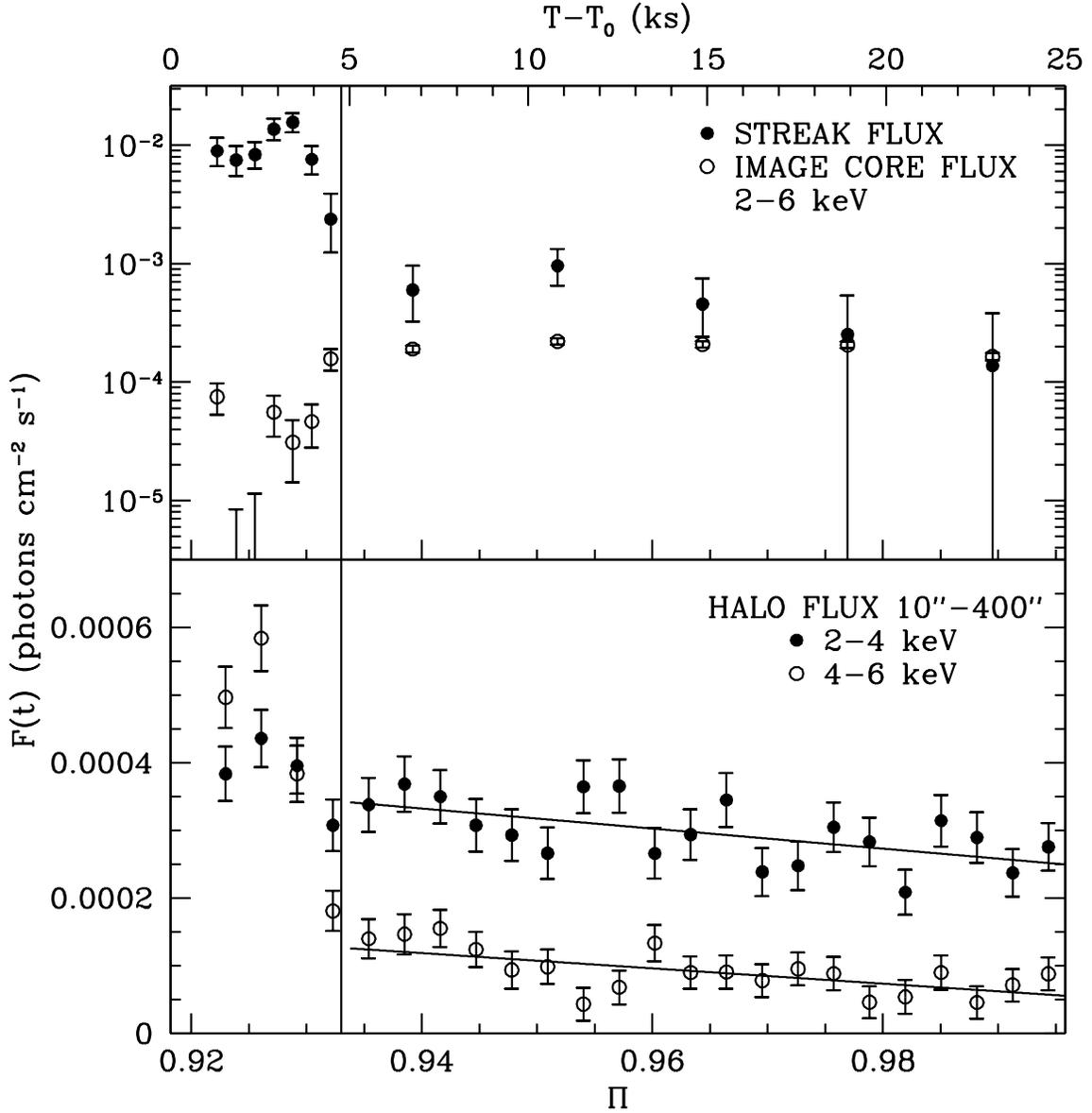} \figcaption[fig3.ps]{Incident fluxes of photons
causing accepted events plotted against time from the start of the
observation and orbital phase. In the upper panel the values shown as
solid circles are derived from events in the transfer streak; the open
circles show values derived from events within $2\arcsec $ of the
image center and depressed by pileup in the CCD chip of the ACIS. The
lower panel shows the background-subtracted halo flux in two energy
ranges. Straight lines were fitted by least squares to the halo
intensities after 5 ksec to show the decay of the halo during the
eclipse.}
\end{figure}

\clearpage
\begin{figure}
\plotone{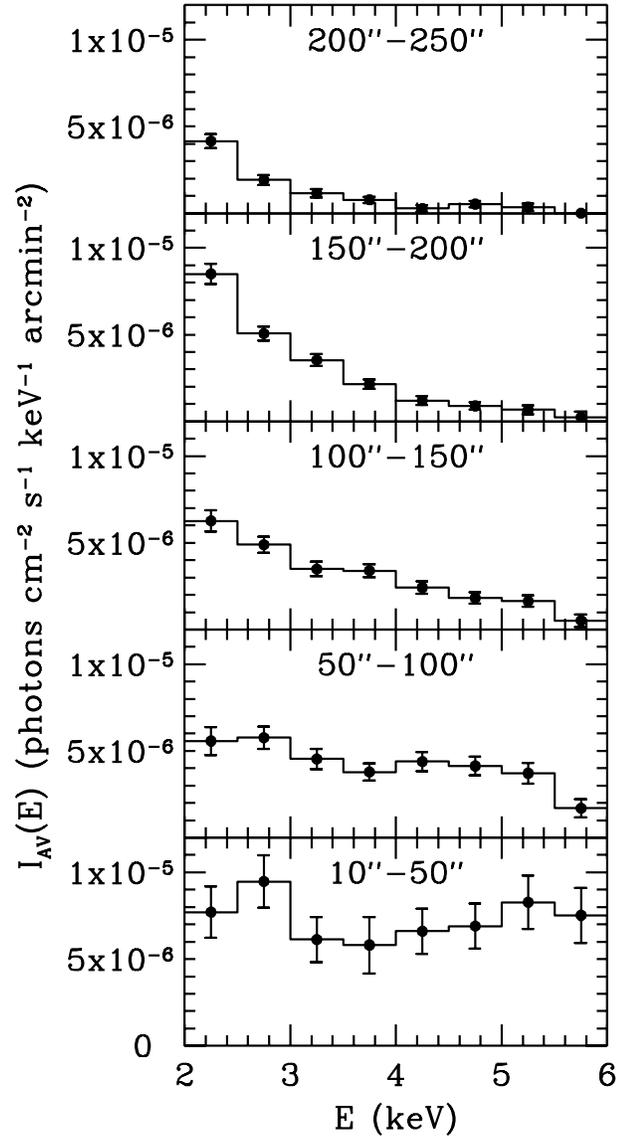} \figcaption[fig4.ps]{Spectra of the
grain-scattered halo in five intervals of angle from the image
center.}
\end{figure}

\clearpage
\begin{figure}
\plotone{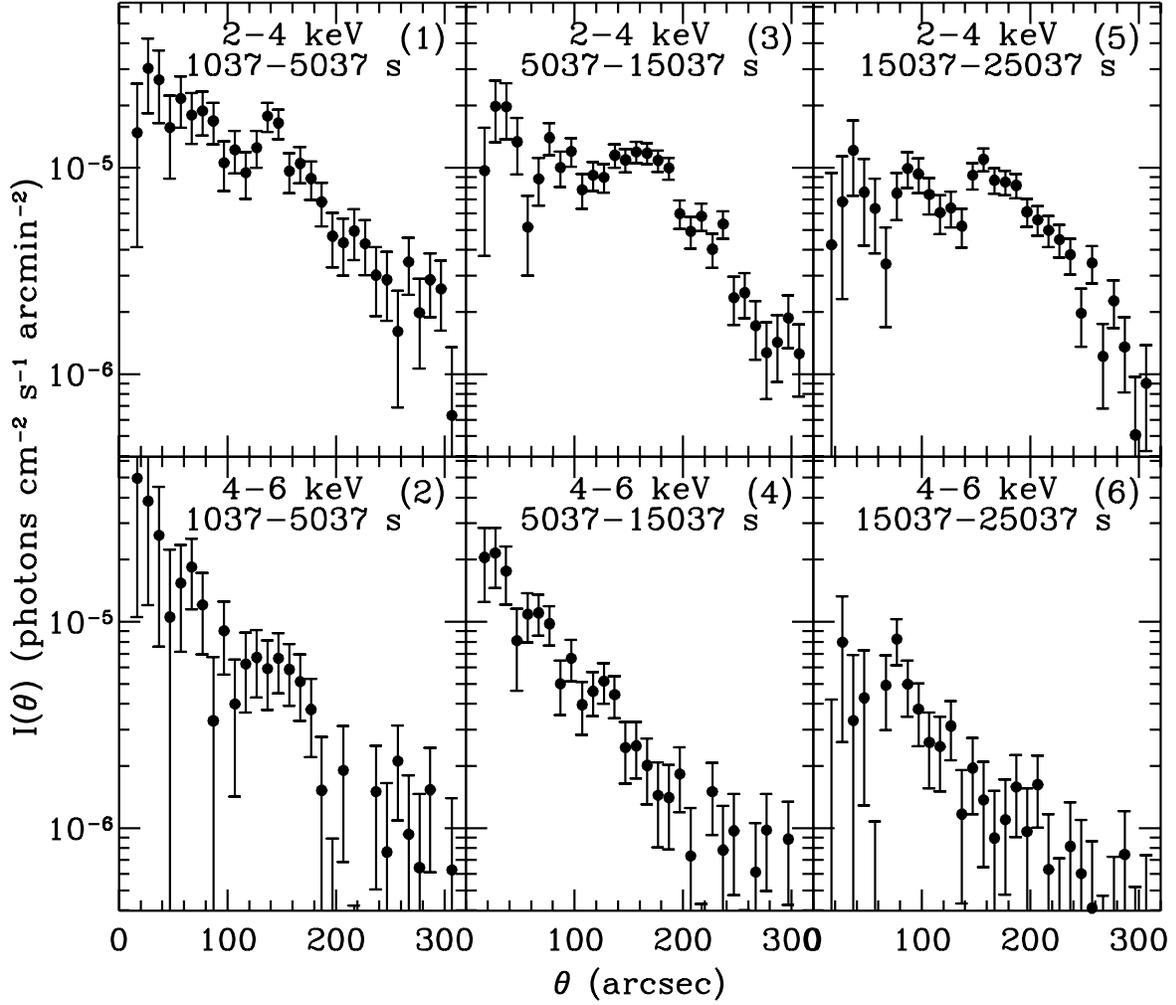} \figcaption[fig5.ps]{Halo profiles in two energy
ranges and three time intervals. The surface brightness is plotted as
intensity against angle from the image center after subtraction of the
PSF of the source image and the background.}
\end{figure}

\clearpage
\begin{figure}
\plotone{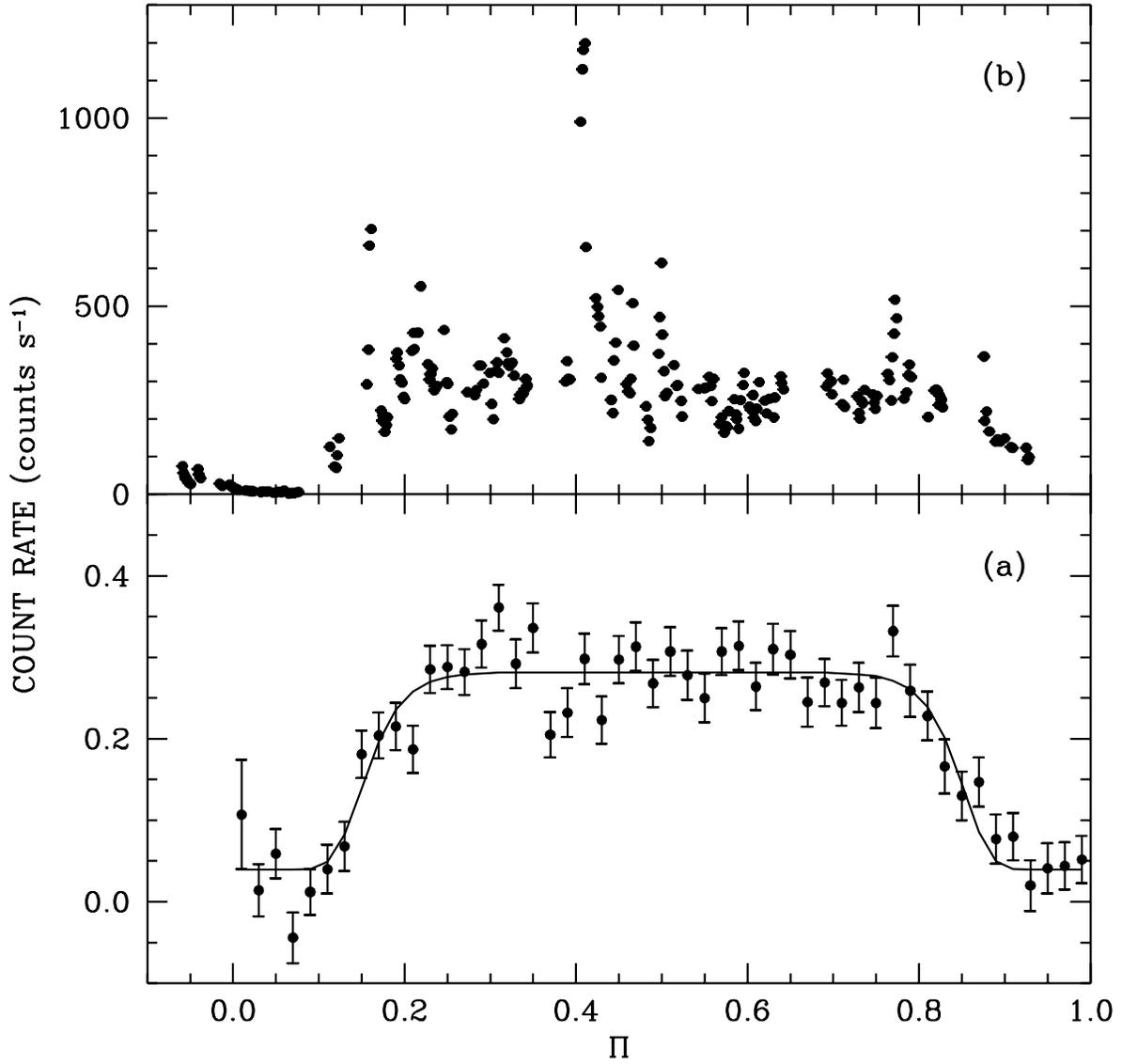} \figcaption[fig6.ps] {(a) RXTE-ASM count rates of
x1538 in fifty equal intervals of orbital phase averaged over 6.91
yr. (b) Average background-subtracted RXTE-PCA count rates of x1538 in
intervals of 529 s plotted against orbital phase as recorded in an
observation during one nearly complete binary orbit.}
\end{figure}

\clearpage
\begin{figure}
\plotone{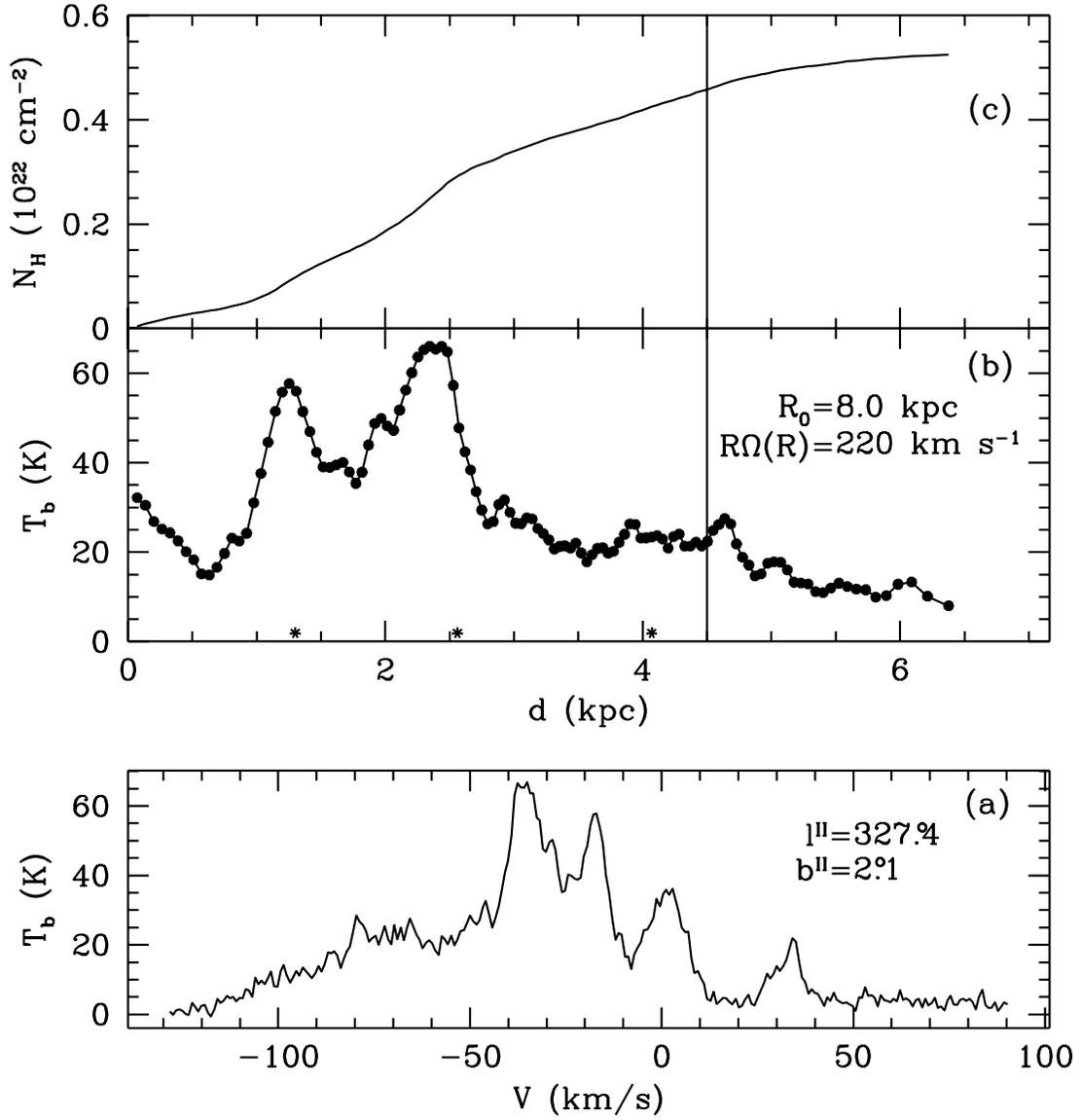} \figcaption[fig7.ps] {(a) Brightness temperature
of 21-cm radiation plotted against radial velocity in the direction of
x1538. (b) Brightness temperature plotted against radial distance
toward x1538. The asterisks mark the distances of the dust clouds and
the vertical line the distance of the source in the simulation model.
(c) Column density of atomic hydrogen plotted against distance toward
x1538.}
\end{figure}

\clearpage
\begin{figure}
\plotone{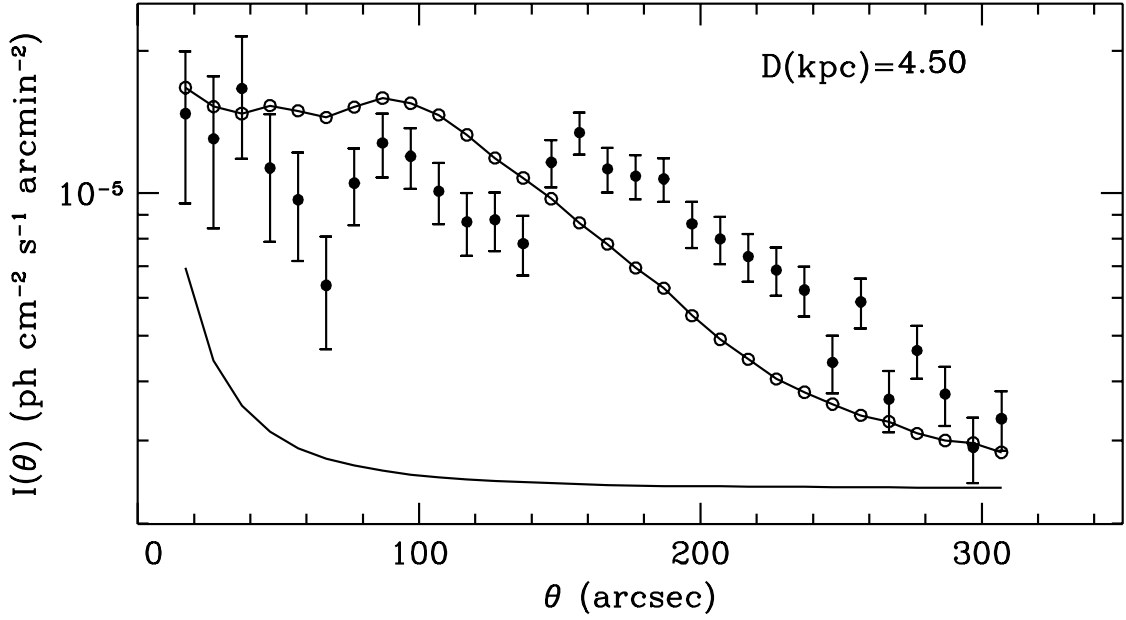} \figcaption[fig8a.ps]{Comparison of the observed
halo profile (solid circles) in the interval from 15037 to 250037 s and
energy range from 2 to 4 keV with the profile (open circles) of a halo
generated from a model of dust distributed with $\tau(d/D)\propto
T_{\rm b}(d)$. The observed profile intensities include background and
PSF intensities. The lower curve indicates the contribution of the
background and PSF to the simulated profile.}
\end{figure}

\clearpage
\begin{figure}
\plotone{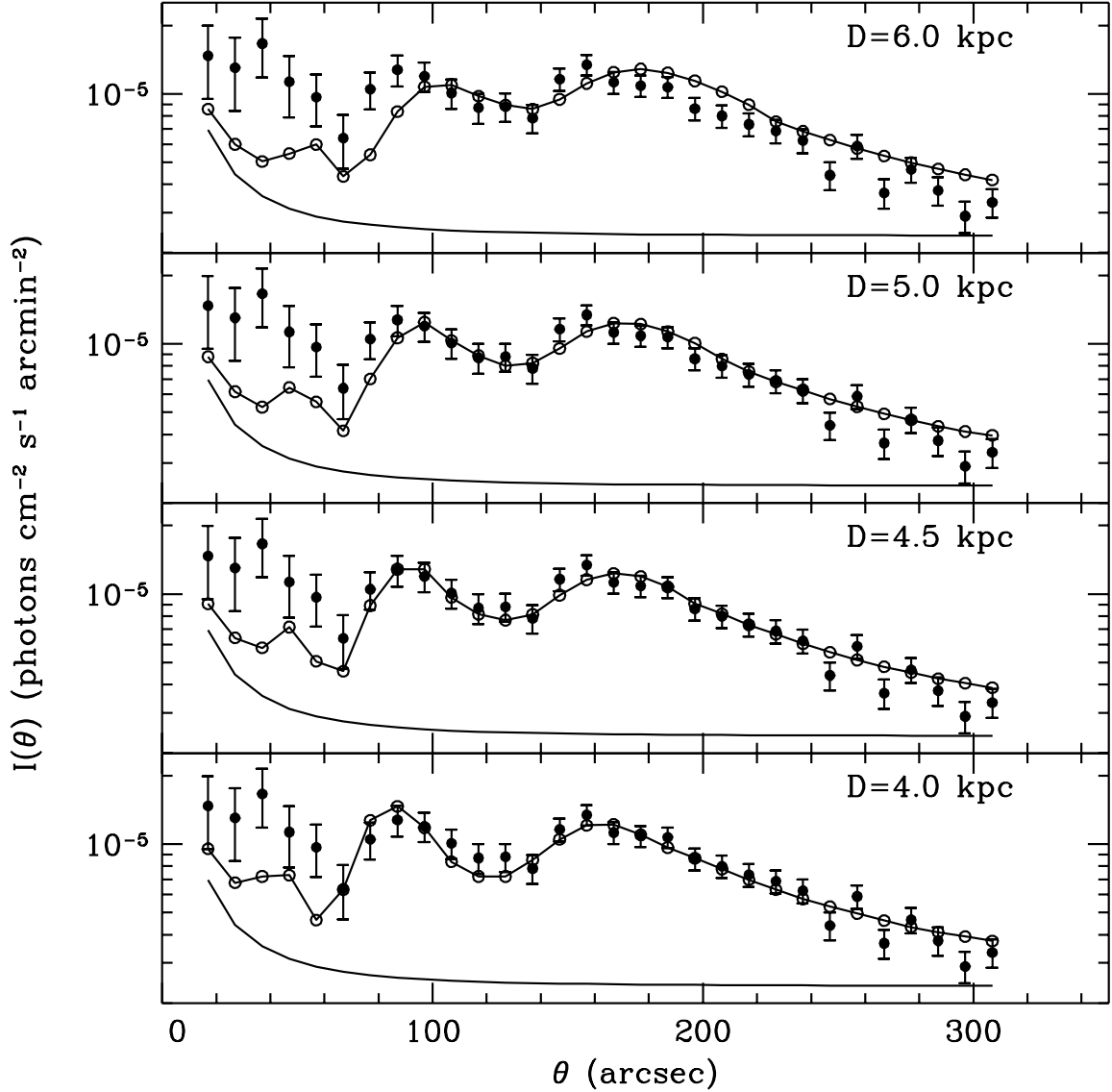} \figcaption[fig8b.ps]{Comparison of the observed
halo profile in the interval from 15037 to 25037 s and energy
range from 2 to 4 keV with fitted simulated halo profiles for
various assumed source distances.  In the models two narrow
dust clouds with adjustable scattering optical depths were placed at
the distances of the two main peaks of brightness temperature in the
radio spectrum of 21-cm radiation in the direction of x1538. The poor
fits for angles less than $60\arcsec$ indicate the need for a third
cloud closer to the source.}
\end{figure}

\clearpage
\begin{figure}
\plotone{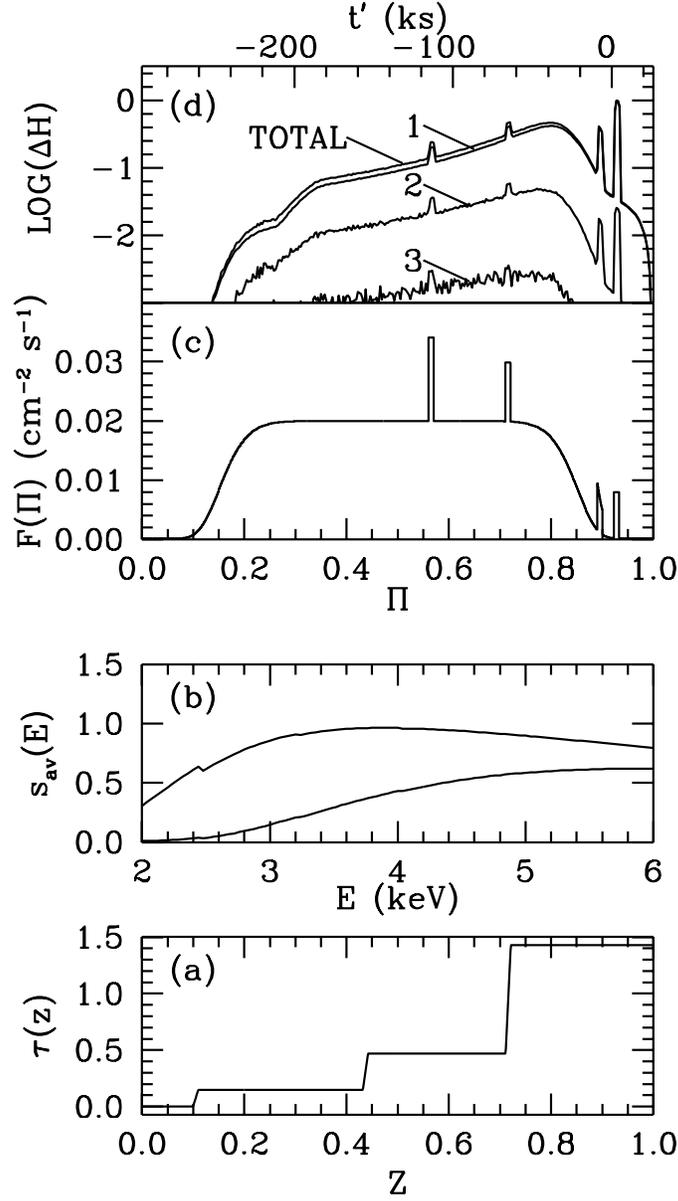} \figcaption[fig9.ps] {Simulation model data. (a)
Scattering optical depth parameter plotted against fractional distance
from the source; (b) Incident average source spectra at two phases of
the binary orbit: top curve at orbital phase 0.5, bottom curve at
orbital phase 0.92 when circumsource photoelectric absorption is
strong; (c) Model incident source flux of 2-6 keV photons plotted
against reduced emission time and orbital phase. The apparent flare
after $t^{\prime}=0$ represents the observed pre-eclipse spectrum; (d)
Relative contributions to the halo flux by photons scattered once,
twice, thrice, and their sum plotted against the reduced emission
time.}

\end{figure}

\clearpage
\begin{figure}
\plotone{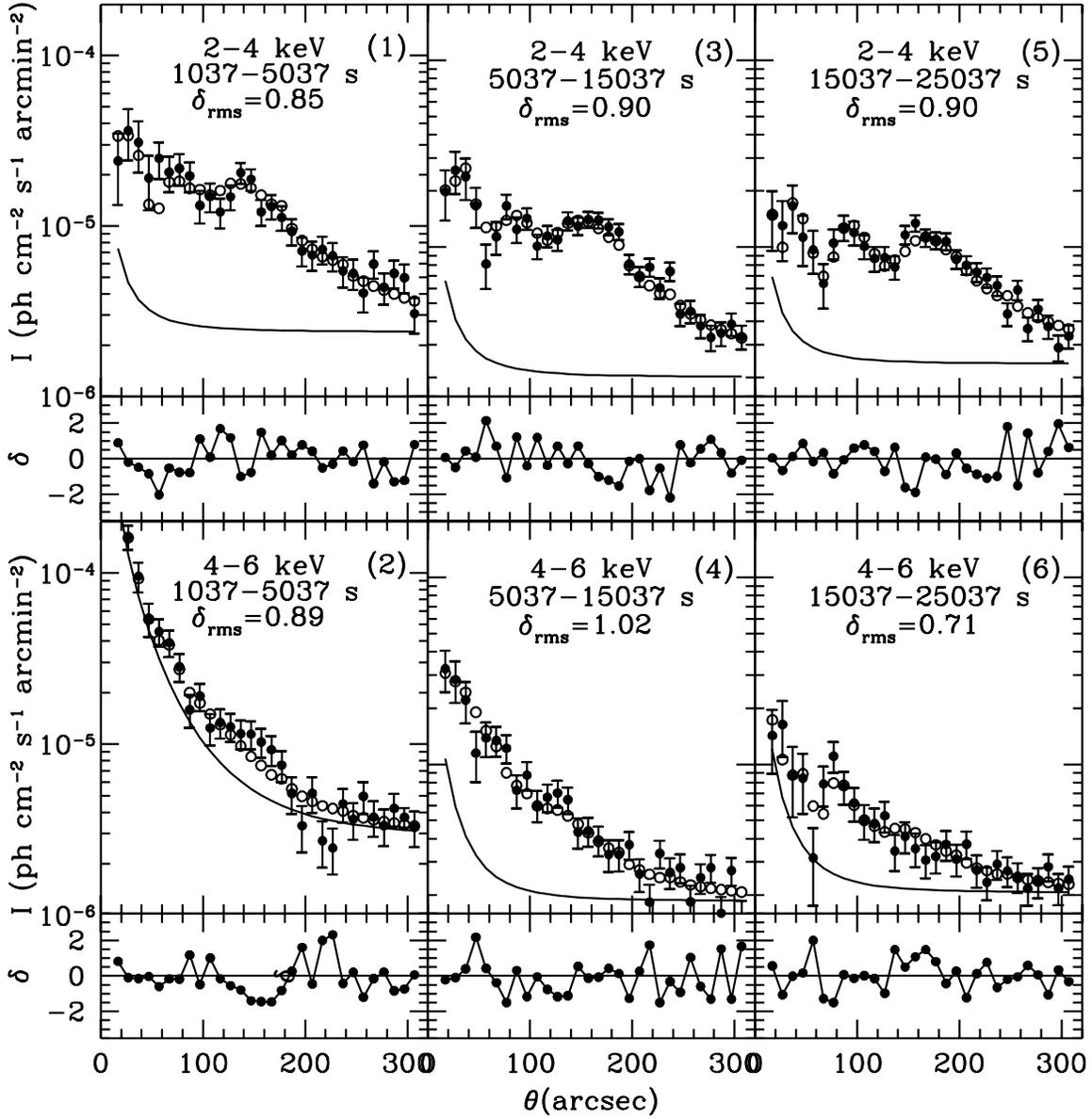} \figcaption[fig10.ps]{Observed halo profiles
without subtraction of background and PSFs of the central image (solid
circles) and fitted simulated halo profiles (open circles). The curves
are the sums of the fitted backgrounds and PSFs that were added to the
simulated halo data. The rms deviation of all six fits is 0.88.}
\end{figure}
 
\clearpage
\begin{figure}  
\plotone{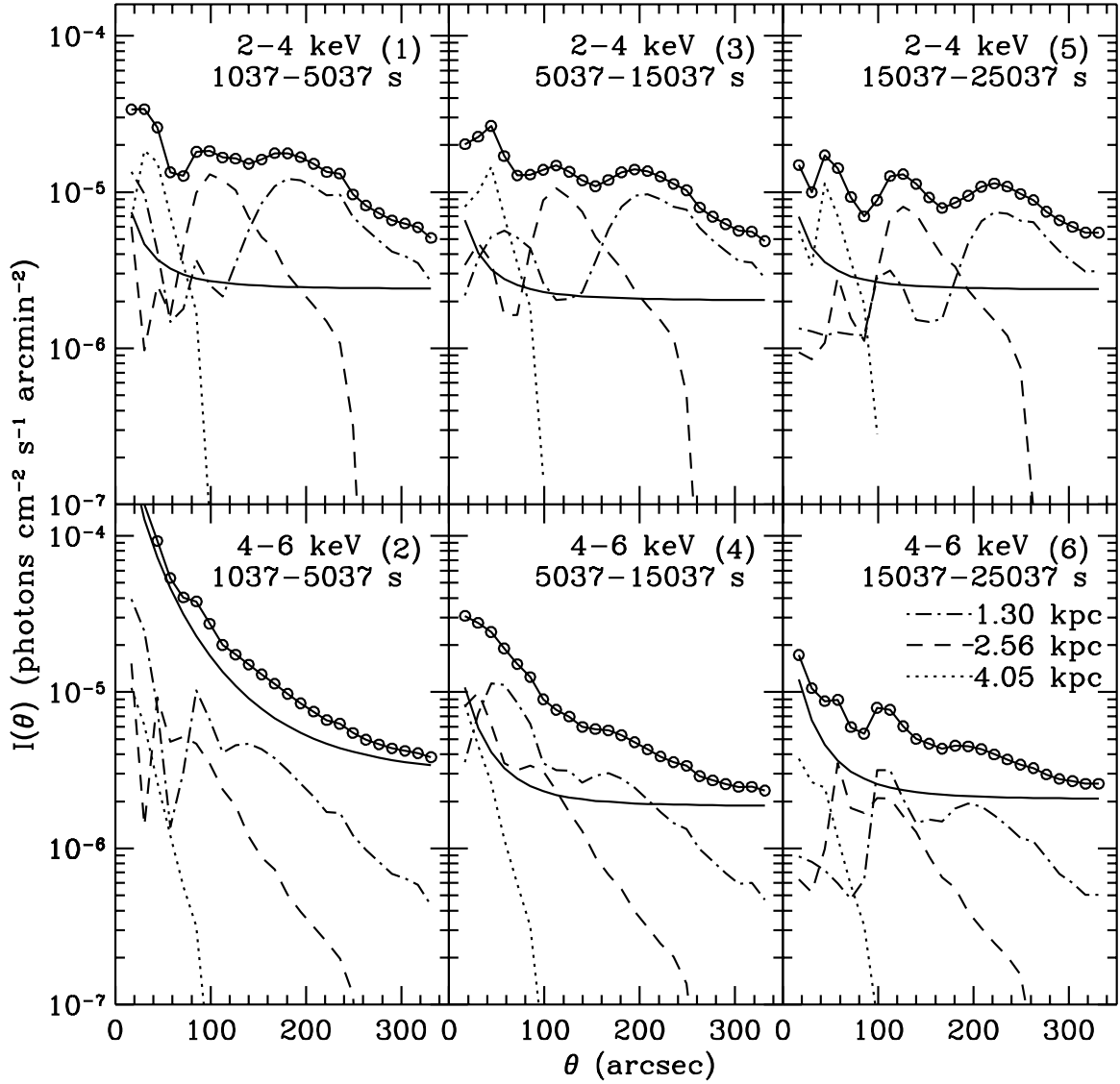} \figcaption[pic/fig11.ps]{Contributions to the
simulated halo profiles from scatterings in the three model dust
clouds. The line types of the contributions are coded according to the
distances of the clouds as indicated in panel (6).}
\end{figure}

\clearpage
\begin{figure} 
\plotone{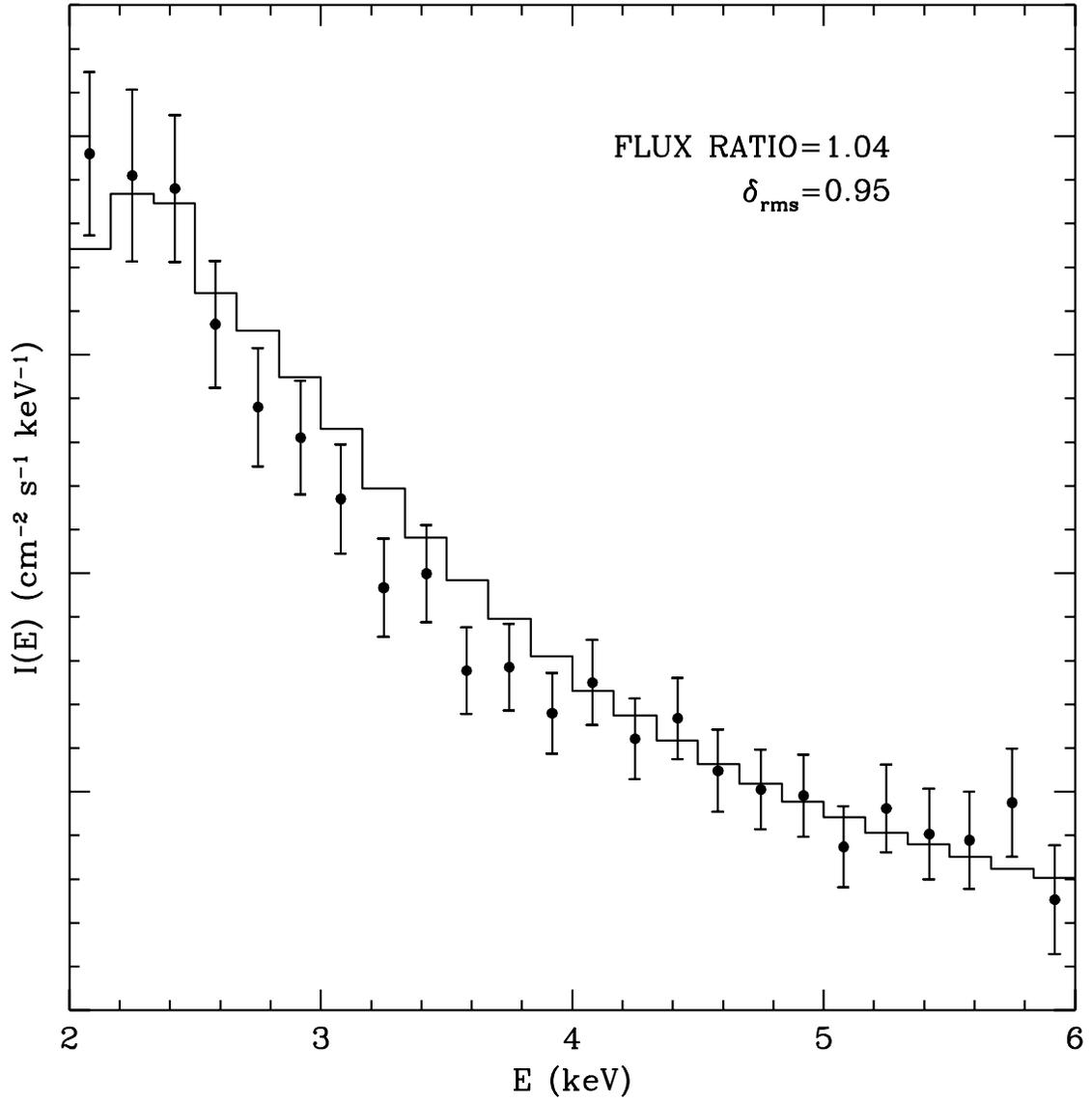} \figcaption[fig12.ps]{Comparison of the spectra
of the observed (solid dots) and simulated (histogram) halo in the
angle range from  30$\arcsec$ to 200$\arcsec$ and time range from 5037
to 25037 s. The rms deviation and the ratio of fluxes in the
simulated and observed halos are shown on the plot.}
\end{figure}

\clearpage
\begin{figure}
\plotone{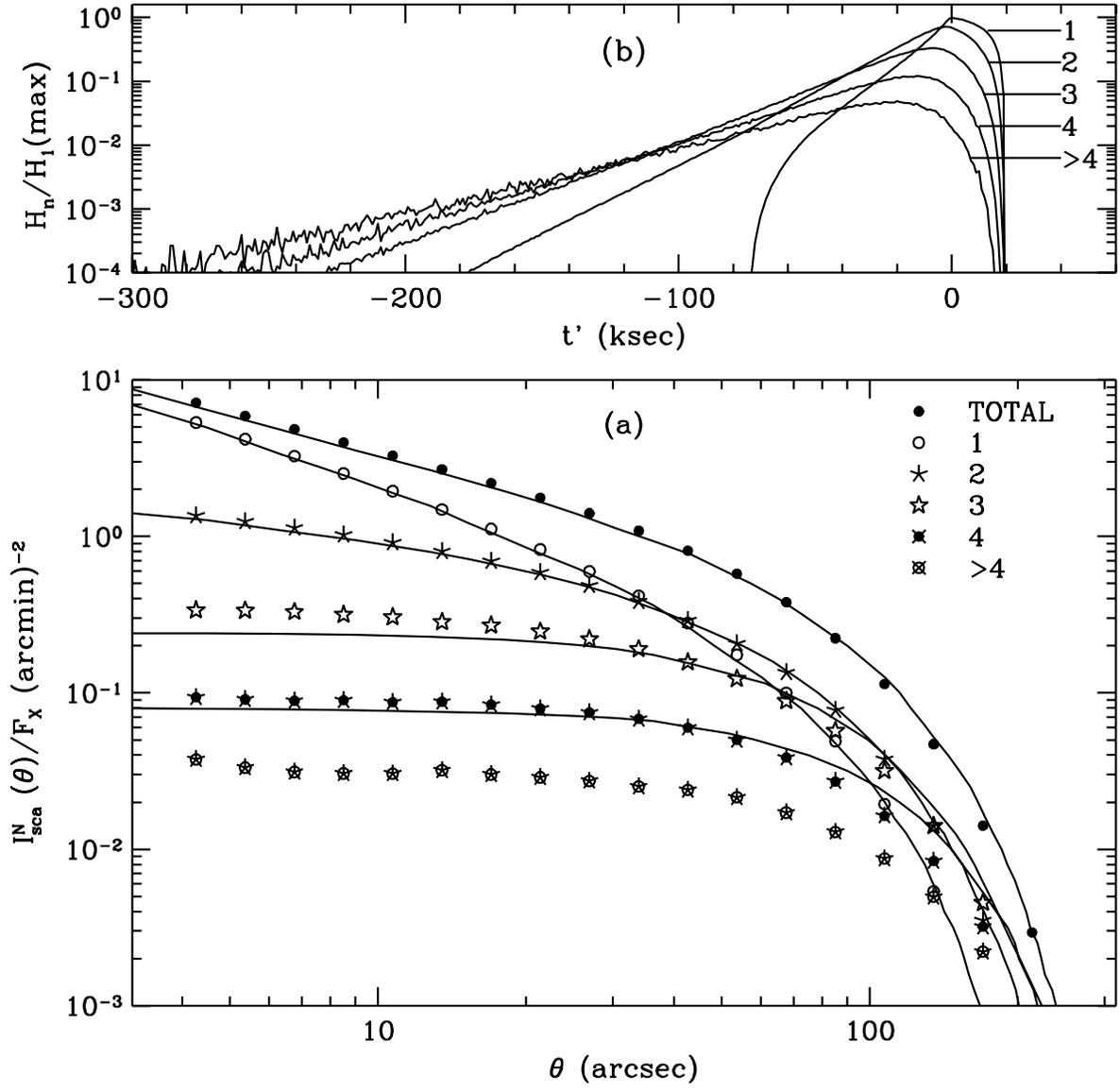} \figcaption[fig14a.ps]{(a) Profiles of a
simulated halo produced by multiple scattering of 1 keV X-rays by
spherical grains with a radius of $1 \mu$ distributed uniformly from a
source at 7 kpc to the observer with $\tau_{scat}=2$. The plotting
symbols are coded according to the number of scatterings as
indicated. The smooth curves are tracings from the plots of analytical
results in Fig. 2 of \citet{mat91}. (b) Relative contributions of
trial photons to the multiply scattered halo components plotted
against their reduced emission times.}
\end{figure}

\clearpage
\begin{figure}
\plotone{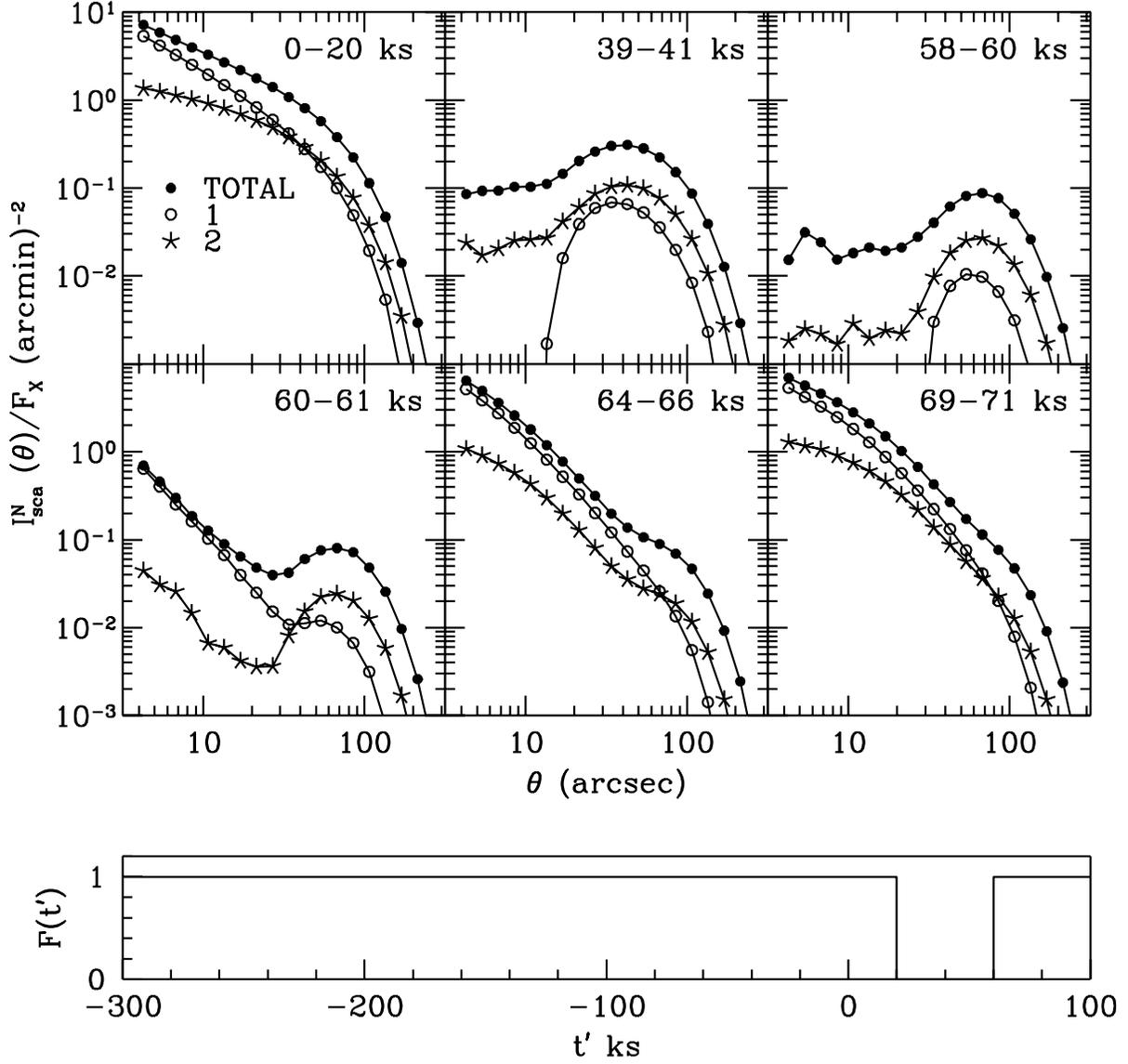} \figcaption[fig15a.ps]{Evolution of the simulated
halo of Fig. 14 before and after the eclipse immersion at 20 ksec.
The flux history is shown in the lower panel. The ``observations''
centered at 40 ksec and 59 ksec show the gradual hollowing out of the
total profile (solid circles) due primarily to the increasing
low-angle cutoff of the single scattered component (open circles), but
partially filled in by the multiply scattered components (asterisks).
The profiles after 60 ksec show the rapid recovery of the profiles
after eclipse emersion.}
\end{figure}

\end{document}